\documentclass[12pt]{article}

\usepackage{amsmath}
\usepackage{wrapfig}
\usepackage{amssymb}
\usepackage{mathrsfs}
\usepackage{color}
\usepackage{graphicx}
\usepackage{mathabx}
\usepackage{dsfont}
\usepackage{float}
\usepackage{epsfig}
\usepackage{lscape}
\usepackage[bottom]{footmisc}
\usepackage{multirow}
\usepackage{booktabs}
\usepackage{caption}
\usepackage{geometry}

\newtheorem{theorem}{Theorem}[section]

\newtheorem{lemma}{Lemma}[section]

\makeatletter

\@addtoreset{equation}{section}
\makeatother

\def\ind{ {{\rm 1}\hskip-2.2pt{\rm l}}}

\renewcommand{\baselinestretch}{1.3}
\usepackage{natbib}

\begin{document}

\title{\bf Non-parametric cure rate estimation under insufficient follow-up using extremes}

\author{
{\large Mikael E\textsc{scobar}-B\textsc{ach}
\footnote{\small{ORSTAT, KU Leuven, Naamsestraat 69, B-3000 Leuven, Belgium. Email: mikael.escobarbach@kuleuven.be and ingrid.vankeilegom@kuleuven.be (corresponding author). Financial support from the European Research Council (2016-2021, Horizon 2020 / ERC grant agreement No.\ 694409) is  gratefully acknowledged.}}} \\ {\it KU Leuven} 
\and 
{\large Ingrid V\textsc{{an} K{eilegom}}} $^*$ \\ {\it KU Leuven}
}

\date{\today}

\maketitle

\begin{abstract}
An important research topic in survival analysis is related to the modelling and estimation of the cure rate, i.e.\ the proportion of subjects that will never experience the event of interest.  However, most estimation methods proposed so far in the literature do not handle the case of insufficient follow-up, that is when the right end point of the support of the censoring time is strictly less than that of the survival time of the susceptible subjects, and consequently these estimators overestimate the cure rate in that case.  We fill this gap by proposing a new estimator of the cure rate that makes use of extrapolation techniques from the area of extreme value theory.  We establish the asymptotic normality of the proposed estimator, and show how the estimator works for small samples by means of a simulation study.  We also illustrate its practical applicability through the analysis of data on the survival of breast cancer patients. 
\end{abstract}

\smallskip
\noindent {\large Key Words:} Cure rate; extreme value theory; right censoring; survival analysis. 

\def\baselinestretch{1.3}
\newpage
\normalsize
\baselineskip 18pt
\setcounter{footnote}{0}
\setcounter{equation}{0}

\section{Introduction}  \label{section_intro}

An emerging research problem in the area of survival analysis is the problem of how to take into account subjects that will never experience the event of interest.  In particular, it is of interest to model and estimate the so-called {\it cure rate} of non-susceptible subjects using flexible models and without making heavy assumptions on the tail of the survival function, since they might not be verified in practice and will lead to biased estimators of the cure rate.  A common assumption under which identifiability and unbiased estimators can be obtained, is the assumption of ``sufficient follow-up'', which means that the censoring time has a larger support than the survival time, and in this case the height of the plateau of the \cite{Kaplan1958} estimator of the survival function estimates consistently the cure rate \citep[see][]{Maller1992}. However, this assumption is often violated in practice, in particular when the duration of the study is short in comparison with the survival times of the non-cured subjects.   In this case, other assumptions are needed in order to identify the unobserved tail of the survival function beyond the last observation.   This paper will offer a solution to this problem using tools from extreme value theory.

The occurrence of cured or non-susceptible subjects is quite common in time-to-event data.  In medical studies, where one is interested in the survival time for a specific disease, some patients might get cured, and the name cure fraction is obviously stemming from this most natural example.  But other examples from diverse areas of applications exist as well, like in economics (duration of unemployment), sociology (age at which someone marries or gets a child), criminology (time before a person released from prison commits a new crime), insurance (time until default), education (time to solve a certain problem), among others.  In all these examples, there is a fraction of the subjects under study that will never experience the event of interest.
   
When the data are exposed to random right censoring, the identification of the cure rate is a complicated problem, since data are then scarce in the right tail of the survival function, and this is exactly  the area where we would like to see sufficient data in order to make the problem identifiable.   A solution to this problem is obtained by assuming that the right end point of the support of the censoring time (denoted by $\tau_c$) is larger than the right end point of the support of the survival time of the susceptible subjects (denoted by $\tau_0$), since in that case we will have data on the full support of the survival time. However, when this assumption is not met, the cure rate cannot be estimated from the data alone as there is insufficient information in the right tail.  This situation has not been well studied in the literature so far, and we aim at filling this gap by using extrapolation techniques from extreme value theory.  Our main idea consists in correcting the Kaplan--Meier (1958) estimator at $\tau_c$ by determining the queue behavior of the distribution of the survival time and extrapolating its value to $\tau_0$. Although our approach is mainly devoted to distributions in the Fr\'echet domain of attraction and thus to cases where $\tau_0=+\infty$, the Weibull max-domain with $\tau_0$ finite is also considered as soon as the latter is known.
   
The literature on nonparametric methods for the estimation of the cure rate is rather scarce compared to the rich literature on the parametric counterpart. Among them, we can cite \cite{Maller1992}, \cite{Peng2000}, \cite{Xu2014}, \cite{Lopez2017}, \cite{Lopez2017b} and \cite{Chown2018}, who consider covariates in the model.  All these papers consider however the case where $\tau_c \ge \tau_0$.   In addition, techniques from extreme value theory have been used in the literature on survival analysis with right censored data when there is no cure fraction. See e.g.\ \cite{Beirlant2001}, \cite{Einmahl2008}, \cite{Beirlant2010}, \cite{Gomes2011}, \cite{Worms2014} and \cite{Stupfler2016}, among others.  However, to the best of our knowledge the problem of cure rate estimation has not been handled so far by using tools from extreme value theory.

The remainder of the paper is organized as follows. In Section \ref{section_model} we describe our estimation method together with the model assumptions. Next, the asymptotic properties of our estimators are presented in Section \ref{section_asymptotic}. In Section \ref{section_simulation} we show how our proposed method works for small samples by means of a simulation study, whereas in Section \ref{section_real} its practical applicability is illustrated through the analysis of data on the survival of breast cancer patients.  Finally, all proofs are collected in Section \ref{section_proofs}.

\section{The estimation method} \label{section_model}


We start with some notations.  The survival time of a subject will be denoted by $T$, and the cure rate is $1-p$, where $p=\mathbb{P}(T<\infty)$. The presence of random right censoring prevents us from observing the survival time for all subjects.  Instead we observe $Y$ and $\delta$, were $Y=\mbox{min}(T,C)$, $\delta=\ind_{\{T \le C\}}$ and the random variable $C$ is the censoring time that is assumed to be finite.   This implies that all cured subjects, i.e.\ those subjects for which $T$ is infinite, are censored, and among the non-cured or susceptibles subjects, some or censored and others are not.  Note that the sub-distribution $F$ of $T$ can be written as
\begin{eqnarray} \label{model}
F(t) = \mathbb{P}(T\leq t) = pF_0(t), 
\end{eqnarray}
where $F_0$ is the distribution of the survival time of the susceptible subjects.  The distribution of the censoring time is denoted by $F_c(t)=\mathbb{P}(C\leq t)$.   Recall also that the right end points of the support of the distributions $F_0$ and $F_c$ are respectively denoted by $\tau_0$ and $\tau_c$.  We will work under minimal conditions on the distribution functions, though we have to impose the usual identification assumption that $T$ and $C$ are independent, which implies that $H(t)=\mathbb{P}(Y \le t)$ satisfies $1-H(t)=(1-F(t))(1-F_c(t))$. 
Finally, suppose we have a sample of independent and identically distributed pairs $\{(Y_i,\delta_i)\}_{1\leq i\leq n}$, having the same distribution as $(Y,\delta)$.

\subsection{Nonparametric estimation under sufficient follow-up} \label{sect2.1}

The nonparametric estimation of the cure rate has been initiated by \cite{Maller1992}.  Their estimator, which we describe below, is consistent under the crucial assumption that the follow-up time is sufficient, which means that 
\begin{eqnarray} \label{followup}
\tau_0 \le \tau_c.
\end{eqnarray}
The estimator is based on the Kaplan-Meier estimator (KME) of the distribution $F$, which is defined as follows. Denote the $i$-th order statistic of $Y_1,\ldots,Y_n$ by $Y_{(i)}$, and denote the corresponding censoring indicator by $\delta_{(i)}$. In the absence of ties, the KME is given by
\begin{eqnarray*}
\widehat{F}_n(t) = 1-\prod_{Y_{(i)}\leq t}\left(1-\dfrac{\delta_{(i)}}{n-1+i}\right), \quad t\in\mathbb{R},
\end{eqnarray*}
where the product over an empty set is defined to be 1. The cure rate $1-p$ is then estimated by the height of the plateau of the KME, or equivalently we estimate $p$ by
\begin{eqnarray*} \label{pestimator}
\widehat{p}_n = \widehat{F}_n(\widehat{\tau}_n),
\end{eqnarray*}
where $\widehat{\tau}_n=Y_{(n)}$ is the largest observed survival time in the sample. The asymptotic consistency of this estimator is stated in the following result, see Theorem 1 in \cite{Maller1992}.  Here, $\tau_H = \inf\{t : H(t)=1\}$ is the right end point of the support of $H$.

\begin{theorem} \label{Maller}
Assume that $0 < p < 1$ and that $F$ is continuous at $\tau_H$ in case $\tau_H<\infty$. Then,  
\begin{eqnarray*}
\widehat{p}_n\rightarrow p\text{ in probability as }n\rightarrow+\infty\quad\text{if and only if }\quad\tau_0\leq\tau_c.
\end{eqnarray*}
\end{theorem}

One of the major implications of this theorem, is that the consistency requires the necessary and sufficient condition that with probability one, no uncured subject can survive longer than the largest possible censoring time. Intuitively, it ensures that we have enough information all over the support of the survival time, such that no observation is almost surely censored.  This condition, which is commonly referred to as the case of \textit{sufficient follow-up}, represents the standard paradigm when it comes to the study of censored data. Nevertheless, it is not always met in practical applications and difficulties might appear for experiments that have a short study duration or a long time to the event of interest. As a matter of fact, the estimator $\widehat{p}_n$ turns out to underestimate $p$ when this condition is not satisfied, i.e.\ when $\tau_c<\tau_0$.  This immediately implies that $\tau_c<+\infty$, whereas $\tau_0$ may be infinite.  In the sequel, we will naturally refer to this situation as the case of \textit{insufficient follow-up}.

\subsection{Extreme value theory}

In order to avoid the condition of sufficient follow-up, we propose to make use of some of the basic concepts from extreme value theory. In particular, we will use the essential idea that queue events can be extrapolated thanks to one single real parameter that characterizes the family of all possible limiting distributions for larger observations in a sample. We thus assume that the large survival times in a sample approximately follow a certain underlying distribution, or in a more formal way, that $F_0$ belongs to the maximum domain of attraction of an extreme value distribution. This means that we assume that there exists a shape parameter $\gamma\in\mathbb{R}$ such that for any $y>0$, 
\begin{eqnarray}
\label{doa}
\lim_{t\rightarrow\tau_0}\dfrac{1-F_0(t+y\ell(t))}{1-F_0(t)}=G_\gamma(y) = 
\left\{
\begin{array}{ll}
(1+\gamma y)^{-1/\gamma} & \text{if }\gamma\neq 0 \\
\exp(-y) & \text{if }\gamma=0,
\end{array}
\right.
\end{eqnarray}
\vspace{.2cm}
where
\begin{eqnarray} \label{doa2}
\ell(t)=\left\{
\begin{array}{ll}
\gamma t & \text{if }\gamma>0 \\
-\gamma(\tau_0-t) & \text{if }\gamma<0 \\
\displaystyle\int_t^{\tau_0} (1-F_0(x)) \, dx/(1-F_0(t)) & \text{if }\gamma=0.
\end{array}
\right.
\end{eqnarray}

The parameter $\gamma$ is called the \textit{extreme value index} and drives the tail behavior of $F_0$.  The case where $\gamma>0$ refers to the Fr\'echet domain of attraction, and it describes distributions with rather heavy tails and $\tau_0=+\infty$.  On the other hand, $\gamma<0$ corresponds to the Weibull domain of attraction, and describes distributions with light tail and $\tau_0<+\infty$.  Finally, when $\gamma=0$ we are in the Gumbel domain of attraction, which describes distributions with tails that have an exponential decay and $\tau_0$ is potentially infinite.

We will focus on distributions $F_0$ with non-zero extreme value index. The reason is that the function $\ell$ defined in (\ref{doa2}) plays a key role in what follows, and for $\gamma=0$ it depends on the function $F_0$ at values up to $\tau_0$, which are unavailable in our framework. Furthermore, without loss of generality we can restrict attention to the case where $\gamma > 0$. Indeed, any random variable $X$ with negative extreme value index $\gamma$ is in the domain of attraction of $G_{-\gamma}$ if and only if 
\begin{eqnarray} \label{psi}
\psi(X) = \dfrac{1}{\tau_0-X},
\end{eqnarray} 
is in the domain of attraction of $G_\gamma$. Hence, the knowledge of $\tau_0$ and a preliminary  transformation by $\psi$ of the observations, allows us to reduce the procedure to the Fr\'echet domain of attraction. Therefore, we will assume from now on and without loss of generality, that survival data with negative extreme value index are initially transformed according to (\ref{psi}).

\subsection{Nonparametric estimation under insufficient follow-up}

As discussed in Section \ref{sect2.1}, the aforementioned estimator $\widehat{p}_n$ is not appropriate in the case of an insufficient follow-up time, since it consistently estimates $F(\tau_c)$, which is in that case smaller than $F(\tau_0)=p$. However, the model in (\ref{doa}) allows us to extrapolate values that are normally out of reach due to the censoring mechanism. The intuitive idea here consists in using the tail behavior of $F_0$ in order to adapt $\widehat{p}_n$, by adding an appropriate correction term that depends on $\widehat\tau_n$, in such a way that the resulting estimator converges to the target value $p$ as $n\rightarrow+\infty$ and $\tau_c\rightarrow\tau_0$. Formally, we replace $t$ by $\tau_c$ and $y$ by $1+\gamma y$ in (\ref{doa}) and obtain
\begin{eqnarray*}
\dfrac{1-F_0(y\tau_c)}{1-F_0(\tau_c)}\simeq y^{-1/\gamma},
\end{eqnarray*}  
where the approximation applies for $\tau_c$ close to $\tau_0$.  Next, a straightforward transformation yields 
\begin{eqnarray*}
\dfrac{F_0(\tau_c)-F_0(y\tau_c)}{1-F_0(\tau_c)}\simeq y^{-1/\gamma}-1,
\end{eqnarray*}  
which by (\ref{model}) is equivalent to
\begin{eqnarray*}
\dfrac{F(\tau_c)-F(y\tau_c)}{p-F(\tau_c)}\simeq y^{-1/\gamma}-1.
\end{eqnarray*} 
Similarly, the use of both $y$ and $y^2$ for $y\neq1$, allows us to extract the particular value of $y^{-1/\gamma}$ free from $p$, that is 
 \begin{eqnarray*}
\dfrac{F(y^2\tau_c)-F(y\tau_c)}{p-F(\tau_c)}\times\dfrac{p-F(\tau_c)}{F(y\tau_c)-F(\tau_c)}=\dfrac{F(y^2\tau_c)-F(y\tau_c)}{F(y\tau_c)-F(\tau_c)}\simeq \dfrac{y^{-2/\gamma}-y^{-1/\gamma}}{y^{-1/\gamma}-1}=y^{-1/\gamma}.
\end{eqnarray*}
Hence, we deduce the following limits : 
\begin{eqnarray}
\label{approximation}
p=\lim_{\tau_c\rightarrow\tau_0} \Big\{F(\tau_c)+\dfrac{F(\tau_c)-F(y\tau_c)}{y^{-1/\gamma}-1}\Big\} \quad \text{and} \quad y^{-1/\gamma}=\lim_{\tau_c\rightarrow\tau_0}\dfrac{F(y^2\tau_c)-F(y\tau_c)}{F(y\tau_c)-F(\tau_c)},
\end{eqnarray}
which leads to our estimator defined by
\begin{eqnarray} \label{defpy}
\widehat{p}_y = \widehat{p}_n+\dfrac{\widehat{F}_n(\widehat{\tau}_n)-\widehat{F}_n(y\widehat{\tau}_n)}{\widehat{y}_\gamma-1}\quad\text{with}\quad\widehat{y}_\gamma:=\dfrac{\widehat{F}_n(y^2\widehat{\tau}_n)-\widehat{F}_n(y\widehat{\tau}_n)}{\widehat{F}_n(y\widehat{\tau}_n)-\widehat{F}_n (\widehat{\tau}_n)},
\end{eqnarray}
for $y\in (0,1)$. Note that $y$ has to be strictly less than 1, since the censoring mechanism exclusively allows us to estimate $F$ at values that do not exceed $\tau_c$. Finally, $\widehat{p}_y$ defines a consistent estimator of the function
\begin{eqnarray*}
p_y(\tau_c) = F(\tau_c)-\dfrac{\left[F(\tau_c)-F(y\tau_c)\right]^2}{F(y^2\tau_c)-2F(y\tau_c)+F(\tau_c)}, 
\end{eqnarray*}
which converges to $p$ as $\tau_c\rightarrow\tau_0$.

\section{Large sample properties} \label{section_asymptotic}

The cornerstone statistic on which the asymptotic properties of our estimator $\widehat{p}_y$ rely, is the Kaplan-Meier estimator $\widehat p_n$.  We therefore recall Theorem 3.14 in \cite{Maller1996}, which states the asymptotic normality of $\widehat{F}_n$ as a process in $D[0,\tau_H]$, endowed with the usual Skorokhod metric.  For this, let $Z=\left\{Z(t)\right\}_{t\in [0,\tau_H]}$ be a stochastic process with independent increments such that $Z(t)$ is normally distributed with mean 0 and variance 
\begin{eqnarray} \label{defv}
v(t)=\int_0^t\dfrac{dF(s)}{(1-F(s))(1-F(s^-))(1-F_c(s^-))},
\end{eqnarray}
and define also the stochastic process $X_n(t)$ on $[0,+\infty)$ given by
\begin{eqnarray} \label{defXn}
X_n(t)=\sqrt{n}\dfrac{1-F(t)}{1-F(t\wedge\widehat{\tau}_n)}\left\lbrace\widehat{F}_n(t)-F(t)\right\rbrace.
\end{eqnarray}
The theorem can then be stated as follows.

\begin{theorem}\label{theo_Fn}
Suppose $F$ is continuous at $\tau_H$ in case $\tau_H<\infty$. Assume also that 
\begin{eqnarray*}
\lim_{t\uparrow\tau_H}(F(\tau_H)-F(t))^2v(t)=0,
\end{eqnarray*}
and that
\begin{eqnarray*}
\lim_{t\uparrow\tau_H}\int_t^{\tau_H}\dfrac{\ind_{\left\{0\leq F_c(s^-)<1\right\}}(1-F(s))}{(1-F_c(s^-))(1-F(s^-))}dF(s)=0.
\end{eqnarray*}
Then, the random variable $\lim_{t\uparrow\tau_H}(1-F(t))Z(t)$ exists and is finite a.s., and, as $n\rightarrow+\infty$, $X_n$ converges weakly in $D[0,\tau_H]$ to the process defined by 
\begin{eqnarray*}
\ind_{\{t\in [0,\tau_H)\}}(1-F(t))Z(t)+\ind_{\{t=\tau_H\}}R,
\end{eqnarray*}
for any $t\geq 0$, where
\begin{eqnarray*}
R=\left\{
\begin{array}{lll}
(1-F(\tau_H))Z(\tau_H),&&\mbox{if }H(\tau_H^-)<1\\
\lim_{t\uparrow\tau_H}(1-F(t))Z(t),&&\mbox{if }H(\tau_H^-)=1.
\end{array}
\right.
\end{eqnarray*}
\end{theorem}

Note that in the definition of the variance function, the continuity of the limiting process is ensured whenever a probability mass is observed at $\tau_H$ for either $F$ or $F_c$. Furthermore, Gill (1994)\nocite{Gill1994} shows that  whenever $F(\tau_H)<1$ and $F_c(\tau_H^{-})<1$, we have on $[0,\tau_H]$,
\begin{eqnarray*}
Z\overset{d}{=}B \circ v,
\end{eqnarray*}  
where $B$ is a standard Brownian motion. This assessment will particularly be helpful for us to derive the asymptotic variance of our estimator. It is also worth to mention that the assumption $F(\tau_H)<1$ is guaranteed by the constraint $\tau_c < \tau_0$. We are now ready to state the asymptotic normality of our estimator $\widehat{p}_y$ defined in (\ref{defpy}). 

\begin{theorem}\label{theo_incidence}
Assume $\tau_c <\tau_0$ and that $F_0$ belongs to a maximum domain of attraction with a non-zero extreme value index, $F_c(\tau_c^{-})<1$ and $F$ is differentiable. Then, under the assumptions of Theorem \ref{theo_Fn} and for any $y\in(0,1)$ such that
\begin{eqnarray*}
\dfrac{F(y^2\tau_c)-F(y\tau_c)}{F(y\tau_c)-F(\tau_c)}\neq 1,
\end{eqnarray*}
we have 
\begin{eqnarray*}
\sqrt{n}\left(\widehat{p}_y-p_y(\tau_c)\right)\overset{d}{\longrightarrow}N(0,\sigma^2_{y,\tau_c})\quad \text{as $n\rightarrow+\infty$},
\end{eqnarray*}
where
\begin{eqnarray*}
\sigma^2_{y,\tau_c}=\sum_{i,j=0}^2 a_i(\tau_c) a_j(\tau_c) \big\{1-F(y^i\tau_c)\big\} \big\{1-F(y^j\tau_c)\big\} v(y^{i\vee j}\tau_c),
\end{eqnarray*}
$a_0(\tau_c) = (1-b_{y,\tau_c})^2$, $a_1(\tau_c) = -2b_{y,\tau_c} (1+b_{y,\tau_c})$, $a_2(\tau_c) = b_{y,\tau_c}^2$, $v(\cdot)$ is defined in (\ref{defv}), and 

\begin{eqnarray*}
b_{y,\tau_c} = \dfrac{F(\tau_c)-F(y\tau_c)}{F(y^2\tau_c)-2F(y\tau_c)+F(\tau_c)}\overset{\tau_c\rightarrow\tau_0}{\longrightarrow}\dfrac{1}{1-y^{-1/\gamma}}.
\end{eqnarray*}
\end{theorem}

\vspace{.2cm}
The proof is given in Section \ref{section_proofs}, and is mainly based on the Skorokhod representation of the process $X_n$ defined in (\ref{defXn}), combined with common Taylor expansions.

\section{Simulations} \label{section_simulation}

In this section we study the finite sample performance of our estimator by means of a simulation study. To this aim, we assume throughout that the censoring time $C$ is uniformly distributed on the interval $[0,\tau_c]$ with probability $1-\varepsilon>0$, and fixed to $\tau_c$ otherwise.   In this way we assure that the condition $F_c(\tau_c^-) < 1$ which was required for the asymptotic theory in the previous section, is satisfied. For the non-cured subjects ($T<\infty$), we consider three models for the survival time $T$ : a standard general Pareto distribution with $\gamma=0.5, 1, 1.5, -0.5, -0.7$ or $-1$ (model 1), a Cauchy distribution with $\gamma=1$ (model 2), or a Beta distribution with positive parameters $(\nu,\mu)$ set to $\nu=1$ and $\gamma=-1/\mu=-0.7$ (model 3).   The proportion of uncured subjects is $p=0.25,0.50$ or 0.75.

We will compare our estimator $\widehat p_y$ and the KME $\widehat p_n$ of $p$ under these different models. To compute $\widehat{p}_y$, a value of $y \in (0,1)$ needs to be chosen.  We choose to work with the $y$-value for which the corresponding estimator $\widehat{p}_y$ is the closest to the average of a bootstrap experiment. For this, consider that any estimator based on a $j$-th uniform resampling of the initial data $(Y_i,\delta_i)$ ($i=1,\ldots,n$) is indexed by $(j)$, $j=1,\ldots,N_b$. Then, the shape parameter $y_*$ is selected as
\begin{eqnarray} \label{ystar}
y_* = \arg\min_{y\in\mathcal{H}}\left|\widehat{p}_y-\dfrac{1}{N_b}\sum_{j=1}^{N_b}\widehat{p}_{y^{(j)}}^{(j)}\right|,\quad\text{with}\quad
y^{(j)}=\sup\left\{y\in\mathcal{H} : \widehat{p}^{(j)}_y>\widehat{p}_n^{(j)}\right\},
\end{eqnarray}
where $\mathcal{H} =\{0.6,0.62,\ldots,0.98\}$ and the supremum over the empty set gives $\widehat{p}_{y^{(j)}}^{(j)} = \widehat{p}_n$.  The motivation for choosing $y$ in this way is that we aim to approximate the distribution $F$ according to $G_\gamma$, which is mostly true in the tail's function. This means that we wish $y$ to be as large as possible, and at the same time, counter-balance the step-function nature of the KME in order to obtain a strictly positive increment $\widehat{F}_n(\widehat{\tau}_n)-\widehat{F}_n(y\widehat{\tau}_n)$.

\begin{figure}[H]
\includegraphics[scale=.29,trim=0cm 0cm 0cm 7cm]{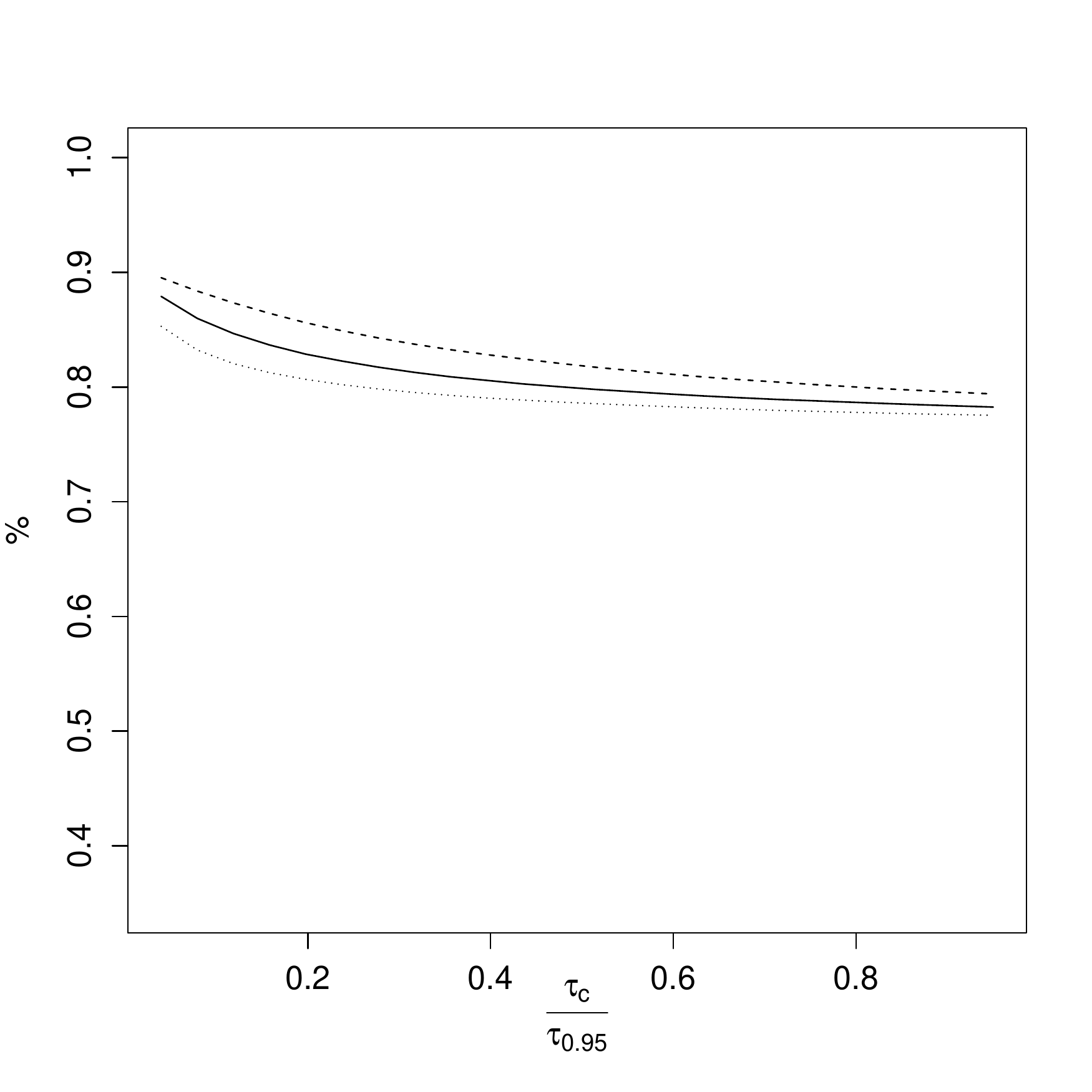}
\includegraphics[scale=.29,page=2,trim=1cm 0cm 1cm 7cm]{figures.pdf}
\includegraphics[scale=.29,page=3,trim=0cm 0cm 0cm 7cm]{figures.pdf}
\caption{Proportion of censoring for $p=0.25$ (left), $p=0.50$ (middle) and $p=0.75$ (right), under model 1 for $\gamma = 0.5$ (dashed curve), 1 (solid curve) and 1.5 (dotted curve).}
\label{fig1}
\end{figure}

\begin{figure}[H]
\includegraphics[scale=.29,page=4]{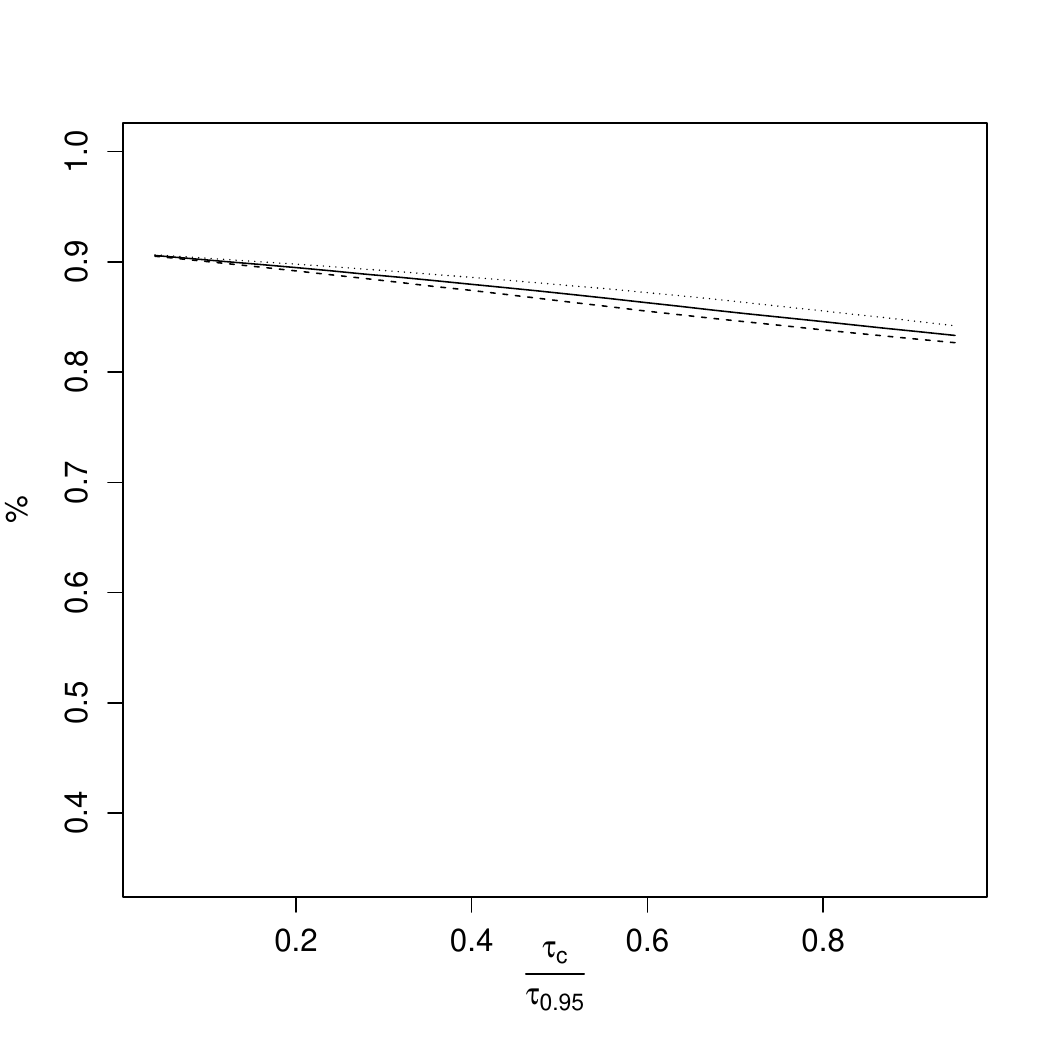}
\includegraphics[scale=.29,page=5,trim=1cm 0cm 1cm 0cm]{figures.pdf}
\includegraphics[scale=.29,page=6]{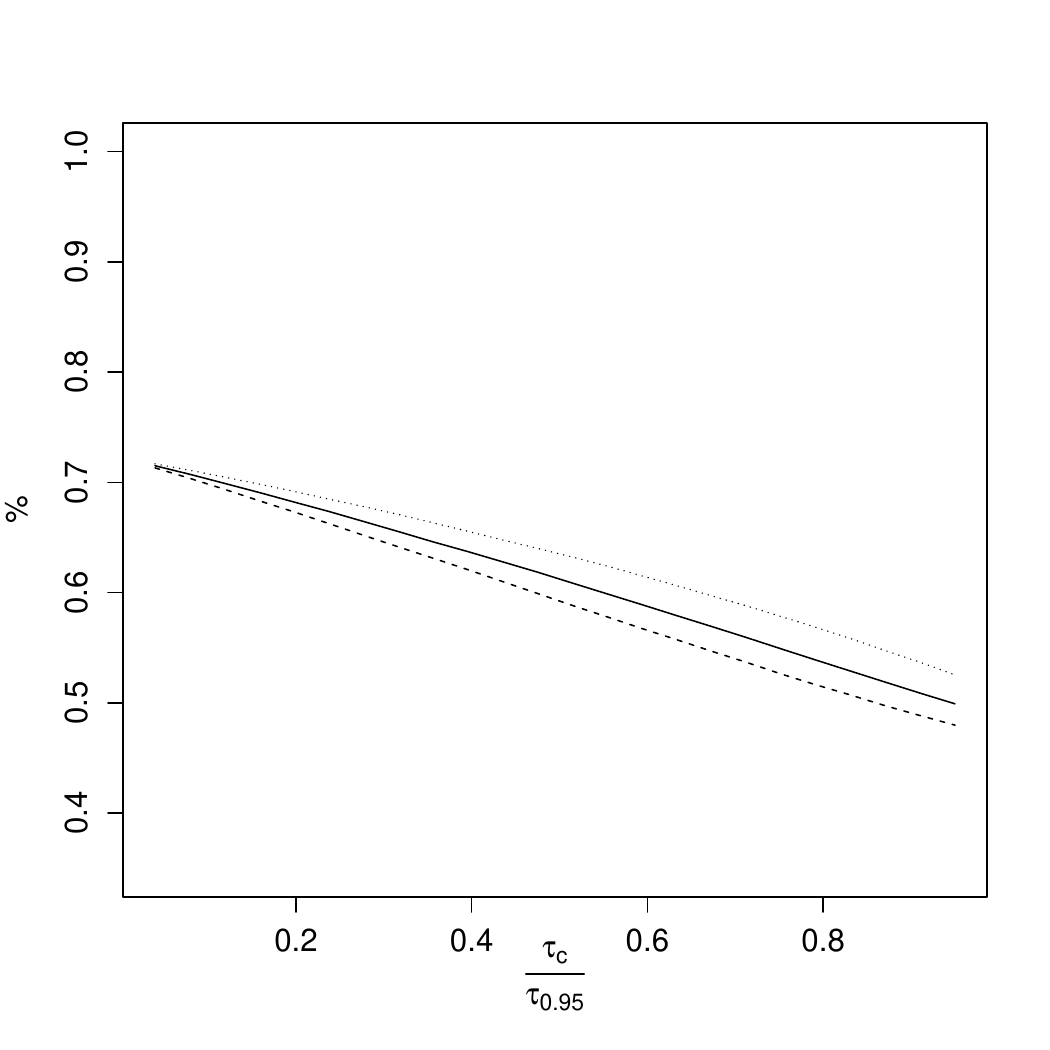}
\caption{Proportion of censoring for $p=0.25$ (left), $p=0.50$ (middle) and $p=0.75$ (right), under model 1 for $\gamma = -0.5$ (dashed curve), $-0.7$ (solid curve) and $-1$ (dotted curve).}
\label{fig2}
\end{figure}

\begin{figure}[H]
\includegraphics[scale=.29,page=7]{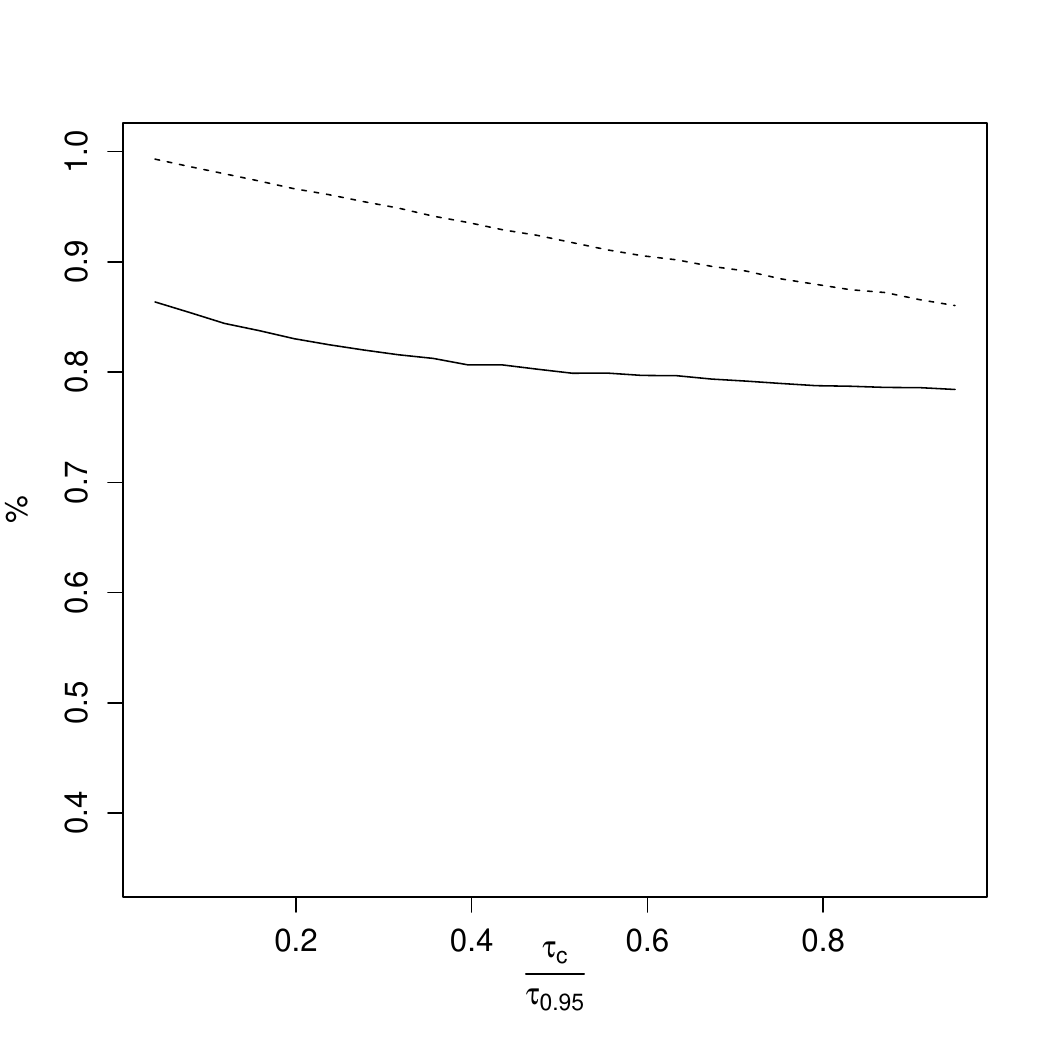}
\includegraphics[scale=.29,page=8,trim=1cm 0cm 1cm 0cm]{figures.pdf}
\includegraphics[scale=.29,page=9]{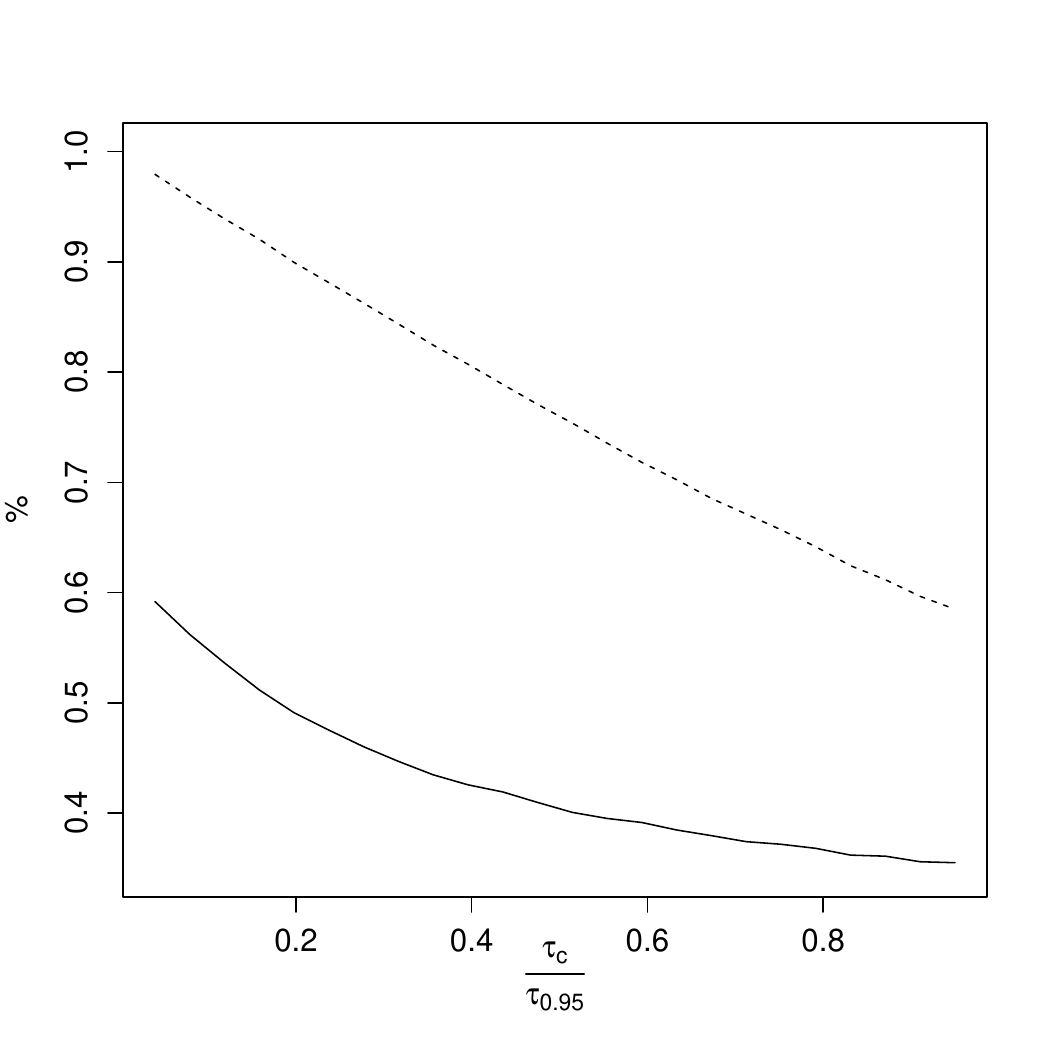}
\caption{Proportion of censoring for $p=0.25$ (left), $p=0.50$ (middle) and $p=0.75$ (right), under model 2 (solid curve) and model 3 (dashed curve).}
\label{fig3}
\end{figure}

Our simulations are based on samples of size $n=1000$ for $N=200$ sample iterations and $N_b=200$ bootstrap iterations, and with $\varepsilon$ fixed to $5\%$. It is also noteworthy to mention that for reasons of homogeneity, the seed of the program generating the samples is fixed.  As a measure of the quality of our estimator, we use the mean squared error given by
\begin{eqnarray*}
MSE(\tau_c) =\dfrac{1}{N}\sum_{j=1}^N\left(\widehat{p}_{y_*,j}-p\right)^2,
\end{eqnarray*} 
where $\widehat{p}_{y_*,j}$ is our estimator obtained for the $j$-th sample iteration. The same is done for the KME $\widehat p_n$.  The results are given in Figures \ref{fig1}--\ref{fig6}, and are represented as a function of the ratio $\tau_c/\tau_{0.95}$ where $\tau_{0.95}$ is the 95th percentile of $F_0$.  The figures are obtained based on a grid of 24 uniformly spaced values of $\tau_c/\tau_{0.95}$ ranging from 0 to 1. In Figures \ref{fig1}--\ref{fig3}, we first show the proportion of censoring for the three models, and next, we present in Figures \ref{fig4}--\ref{fig6} the average and the MSE of our estimator $\widehat p_{y_*}$ and of the KME $\widehat p_n$ under the different model setups.   

\newpage

\begin{figure}[H]
\includegraphics[scale=.4,trim = 0cm 1cm 0cm 2cm,page=10]{figures.pdf}
\includegraphics[scale=.4,page=2,trim = 0cm 1cm 0cm 2cm,page=11]{figures.pdf}
\includegraphics[scale=.4,trim = 0cm 1cm 0cm 0cm,page=13]{figures.pdf}\hfill
\includegraphics[scale=.4,page=2,trim = 0cm 1cm 0cm 0cm,page=14]{figures.pdf}\hfill
\includegraphics[scale=.4,trim = 0cm 1cm 0cm 0cm,page=16]{figures.pdf}\hfill
\includegraphics[scale=.4,page=2,trim = 0cm 1cm 0cm 0cm,page=17]{figures.pdf}
\caption{Average (left) and MSE (right) of $\widehat{p}_{y_*}$ (curve) and $\widehat p_n$ (curve + circle) for model 1 with $p=0.25,0.50,0.75$ from the top to the bottom and $\gamma=0.5$ (dashed), $\gamma=1$ (solid) and $\gamma=1.5$ (dotted). The true value of $p$ is represented by the horizontal line.}
\label{fig4}
\end{figure}

\begin{figure}[H]
\includegraphics[scale=.4,trim = 0cm 1cm 0cm 2cm,page=19]{figures.pdf}
\includegraphics[scale=.4,page=2,trim = 0cm 1cm 0cm 2cm,page=20]{figures.pdf}
\includegraphics[scale=.4,trim = 0cm 1cm 0cm 0cm,page=22]{figures.pdf}\hfill
\includegraphics[scale=.4,page=2,trim = 0cm 1cm 0cm 0cm,page=23]{figures.pdf}\hfill
\includegraphics[scale=.4,trim = 0cm 1cm 0cm 0cm,page=25]{figures.pdf}\hfill
\includegraphics[scale=.4,page=2,trim = 0cm 1cm 0cm 0cm,page=26]{figures.pdf}
\caption{Average (left) and MSE (right) of $\widehat{p}_{y_*}$ (curve) and $\widehat p_n$ (curve + circle) for model 1 with $p=0.25,0.50,0.75$ from the top to the bottom and with $\gamma=-0.5$ (dashed), $\gamma=-0.7$ (solid) and $\gamma=-1$ (dotted). The true value of $p$ is represented by the horizontal line.}
\label{fig5}
\end{figure}

\begin{figure}[H]
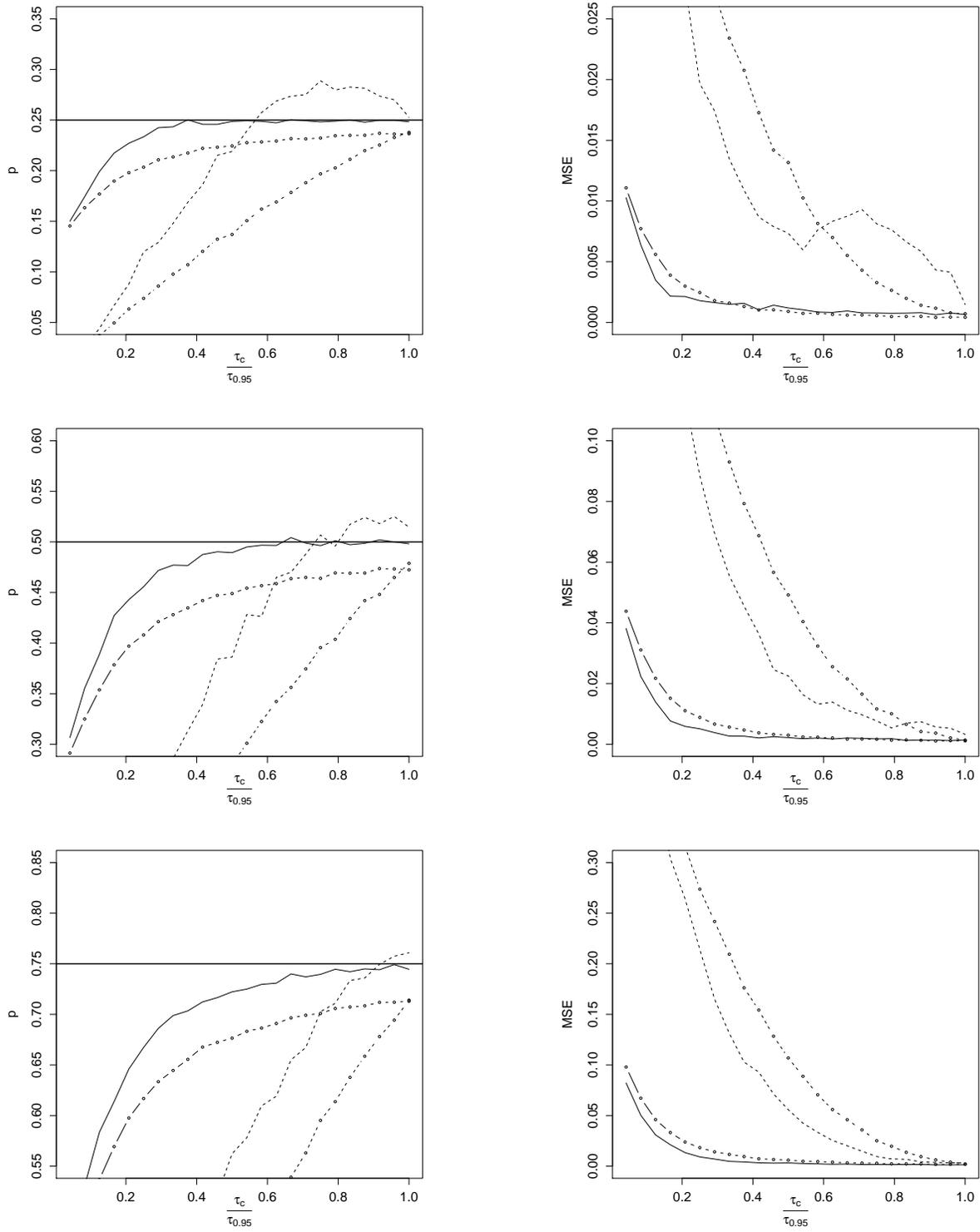

\includegraphics[scale=.4,trim = 0cm 1cm 0cm 2cm,page=28]{figures.pdf}\hfill
\includegraphics[scale=.4,page=2,trim = 0cm 1cm 0cm 2cm,page=29]{figures.pdf}\hfill
\includegraphics[scale=.4,trim = 0cm 1cm 0cm 0cm,page=31]{figures.pdf}\hfill
\includegraphics[scale=.4,page=2,trim = 0cm 1cm 0cm 0cm,page=32]{figures.pdf}\hfill
\includegraphics[scale=.4,trim = 0cm 1cm 0cm 0cm,page=34]{figures.pdf}\hfill
\includegraphics[scale=.4,page=2,trim = 0cm 1cm 0cm 0cm,page=35]{figures.pdf}\hfill
\caption{Average (left) and MSE (right) of $\widehat{p}_{y_*}$ (curve) and $\widehat p_n$ (curve + circle) for model 2 (solid) and model 3 (dashed) with $p=0.25,0.50,0.75$ from the top to the bottom. The true value of $p$ is represented by the horizontal line.}
\label{fig6}
\end{figure}

Based on these simulations, we can draw the following conclusions. 
From Figures \ref{fig1}--\ref{fig3} we see that the proportion of censoring, which by construction is at least equal to the cure rate $1-p$, lies for all models between 35$\%$ and 90$\%$.  However, models with $\gamma<0$ suffer much more from the censoring than models with $\gamma>0$.
The KME $\widehat{p}_n$ never reaches $p$ even when $\tau_c$ approaches $\tau_0$. This is theoretically expected since the estimator $\widehat{p}_n$ consistently estimates $F(\tau_c)$, which is strictly smaller than $p$ when $\tau_c<\tau_0$. 
We also see that the estimation accuracy clearly improves when $p$ increases, which can be explained by the fact that larger values of $p$ lead to more uncensored observations, while the number of cured subjects decreases. However, the minimal value of $\tau_c/\tau_{0.95}$ for which good results are obtained also increases along with $p$.
Another interesting feature is that the estimators $\widehat p_{y_*}$ and $\widehat p_n$ provide more satisfactory results for $\gamma>0$ than for $\gamma<0$, for a wider range of values of $\tau_c$. This can probably be explained by the higher level of censoring of which the data suffer in the Weibull domain of attraction compared to the Fr\'echet domain.
Also, observe that our estimator $\widehat p_{y_*}$ always outperforms the KME $\widehat p_n$, and manages to reach the target value of $p$ as soon as $\tau_c/\tau_{0.95}$ is large enough. In the worst case, a level of $80\%$ is necessary but in most of the cases it works well starting from around $60\%$, with a minimum approximately at $40\%$.  
Finally, in terms of the MSE, our estimator generally shows lower or similar MSE curves than the KME for a wide range of $\tau_c/\tau_{0.95}$, and the two curves tend to merge for $\tau_c$ close to $\tau_0$. They are in general decreasing, except for our estimator in some cases, and reach their minimal values as $\tau_c \to \tau_0$.    

Overall we can conclude that the use of the KME $\widehat p_n$ is clearly not appropriate when there is an insufficient follow-up. This simulation study shows that our approach leads to a particularly efficient bias reduction against the usual Kaplan Meier estimator, with the nice feature of being also less costly in terms of MSE. The effectiveness of the proposed estimator $\widehat p_{y_*}$ depends on the value of the ratio $\tau_c/\tau_{0.95}$, with a threshold depending on the value of the extreme value index and the cure rate.

\section{Real data application} \label{section_real}

\subsection{Background}

Due to a limited recording time, it is often difficult to determine survival rates for slowly proliferating tumors. Oncologists therefore tend to prefer the use of the 5 or 10 years survival rates instead of the actual cure rates. As a matter of fact, \cite{Tai2005} particularly conclude that the follow-up is insufficient for two slowly proliferating cancers, namely thyroid and breast cancer, with data sources based on the Surveillance, Epidemiology and End Results (SEER) database that regroups clinical, pathological and demographic information on cancer patients since 1973. In their study, they particularly identify the threshold of the year at which the cure rate can be correctly estimated. It turns out that the stages III+IV of breast cancer admit a threshold at 20.7 years while the actual records in the SEER data base are limited to 11 years. This means that this situation perfectly fits our context of insufficient follow-up and thus offers an interesting application of our method.

Additionally, we also propose to further inspect the cure rate by racial/ethnic origins and cancer molecular subtypes. Indeed, studies from the past decades have emphasized the disparities between black and white women regarding the incidence, mortality and survival rate of breast cancer. We can cite among others, \cite{Basset1986}, \cite{Eley1994} and \cite{Hunt2014}. The most recent investigations suggest a convergence of the incidence rates among the racial/ethnic groups, while the mortality rate inequalities keep widening among the black and white populations (see \cite{DeSantis2016} and \cite{DeSantis2017}).  

\subsection{Data analysis}

\begin{figure}[H]
\centering
\includegraphics[scale=.35,page=37]{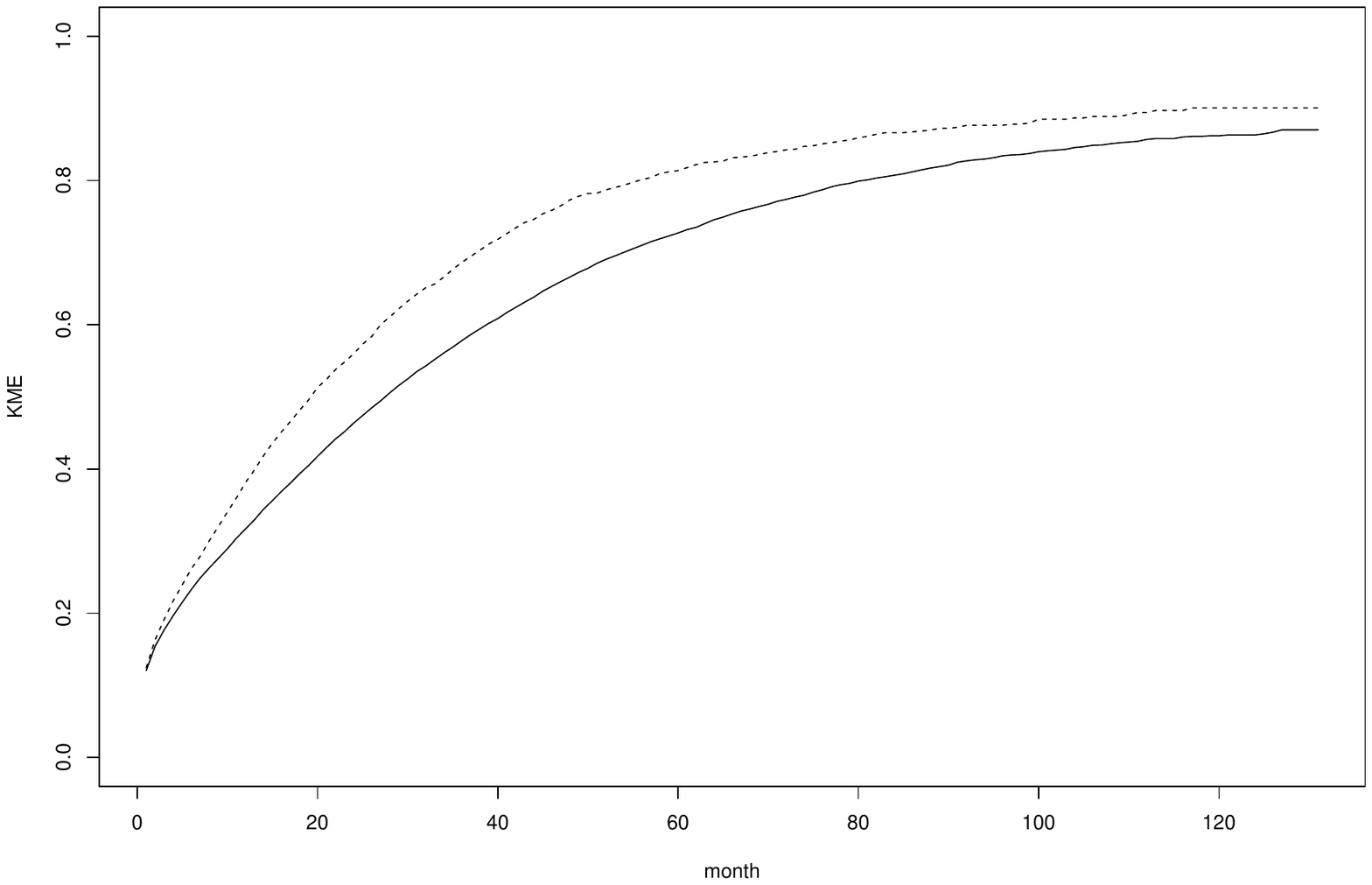}
\caption{Kaplan-Meier estimators for the breast cancer data at stage IV with known subtype for women across the U.S. among the white (full curve) and black (dashed curve) population.}
\label{figure_KME}
\end{figure}

Like in previous studies, we base our study on the survival times of breast cancer at stage IV for 26\,301 non-Hispanic black and white women from the SEER database. The data range from 1975 to 2016 and cover $28\%$ of the U.S. population.  The distribution of the survival time of the susceptibles must be in the domain of attraction of $G_\gamma$ for some $\gamma\in\mathbb{R}^+$. To the best of our knowledge no method has been developed so far for testing if right censored observations belong to a maximum domain of attraction. However, this assumption is quite common and hardly imposes any restriction on the applicability of our method. Hence, we justify our choice by looking at the curves of the KME in Figure \ref{figure_KME}.  The concavity of these curves was also observed for the curves in the simulation study in Section \ref{section_simulation}. This suggests that the $\gamma$ parameter for the breast cancer data at stage IV is indeed positive. 

Breast cancer can be categorized in up to 4 different molecular subtypes, and the classification depends on the presence of receptors with respect to oestrogen and/or progesterone hormones (HR+/HR-) and human epidermal growth factor 2 (HER2+/HER2-). Essentially, the subgroups are HR+/HER2-, HR+/HER2+, HR-/HER2+ and TNBC for the triple negative receptors expression.  A final subgroup is the group of patients that cannot be associated to one of these four subtypes.  This classification forms the basis of our comparison between black and white women.   More details about the distribution of patients among the subgroups and about the proportion of censoring in each subgroup can be found in Table \ref{table}.

\begin{center}
\begin{tabular}{c|rc|rc|}
\cline{2-5}
& \multicolumn{2}{c|}{White} & \multicolumn{2}{c|}{Black} \\ 
\hline
\multicolumn{1}{|c|}{Subtype} & $n$ & censoring ($\%$) & $n$ & censoring ($\%$) \\ 
\hline
\multicolumn{1}{|c|}{All} & 21555 & 60 & 4746 & 65 \\
\hline
\multicolumn{1}{|c|}{HR+/HER2-} & 5491 & 39 & 1033 & 42 \\
\multicolumn{1}{|c|}{HR+/HER2+} & 1464 & 31 & 350 & 36 \\
\multicolumn{1}{|c|}{TNBC} & 1117 & 64 & 433 & 65 \\
\multicolumn{1}{|c|}{HR-/HER2+} & 768 & 39 & 199 & 47\\
\hline
\multicolumn{1}{|c|}{Others} & 12715 & 74 & 2731 & 78\\
\hline
\end{tabular}
\captionof{table}{Sample size and proportion of censoring for the different subgroups of the breast cancer data.}
\label{table}
\end{center}

\begin{figure}[H]
\centering
\includegraphics[scale=.5,page=38]{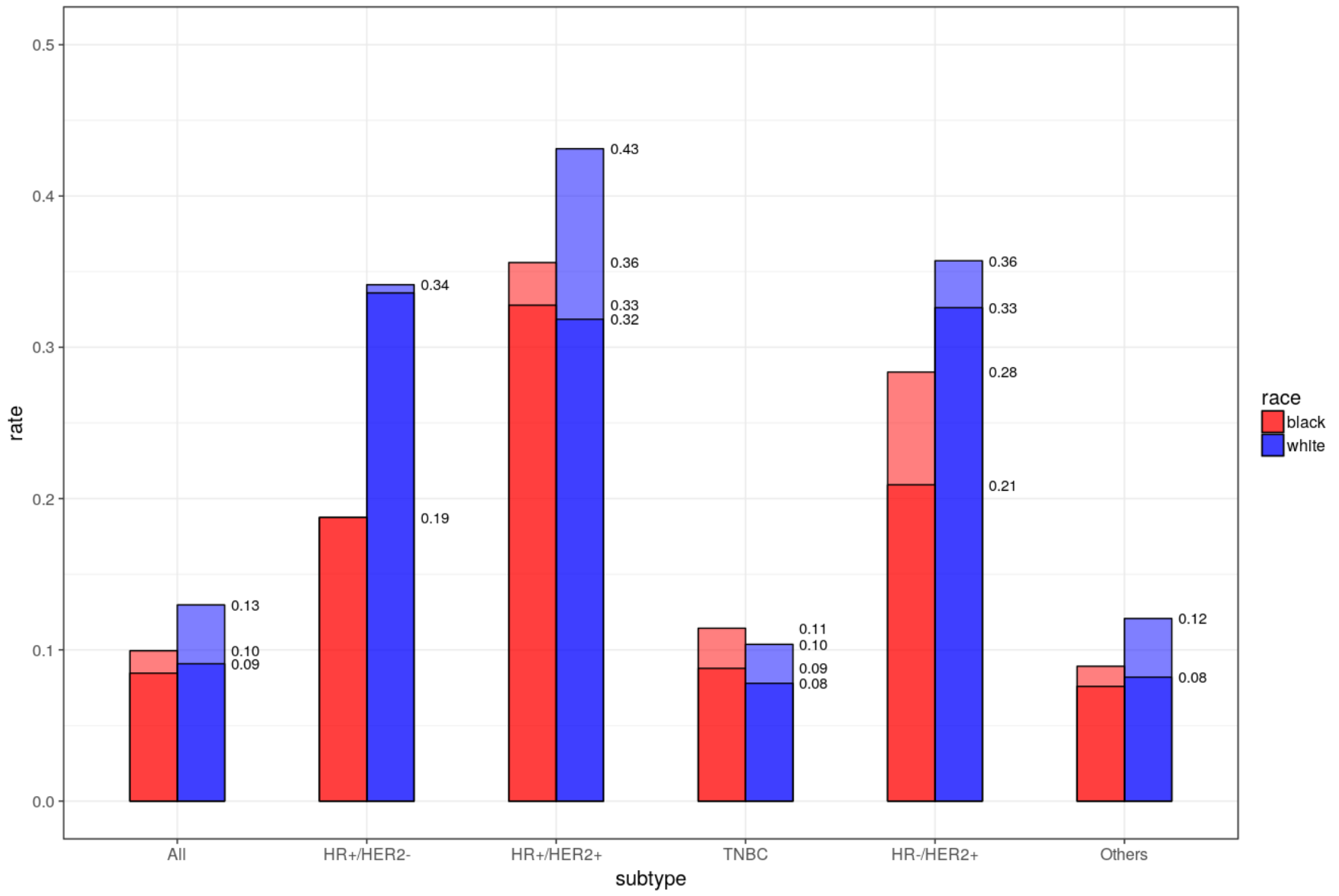}
\caption{Bar-plot representation of the two estimators of the cure rate $1-p$ : the KME $1-\widehat p_n$ (light) and the proposed estimator $1-\widehat{p}_{y_*}$ (dark) for the breast cancer data among the black (red) and white (blue) women.}
\label{figure_barplots}
\end{figure}

In Figure \ref{figure_barplots} we represent the estimation results of the cure rates for the complete dataset and also for each subtype separately.  We use both the KME $1-\widehat p_n$ and the proposed estimator $1-\widehat p_{y_*}$, with $y_*$ as in (\ref{ystar}).  As expected, the KME is always higher than or equal to our estimator, but there are some important differences between the subgroups. 
For the complete data set the cure rate drops by about $10\%$ for all women, and a similar behavior is observed for the subgroups TNBC and `Others' at slightly lower levels. 
For the HR+/HER2- subgroup we see that our estimator and the KME are almost equal for black and for white women.
The HR+/HER2+ and HR-/HER2+ subgroups show important discrepancies between the KME and our estimator, especially for white women in the HR+/HER2+ subgroup (reduction from $43\%$ to $32\%$) and for black women in the HR-/HER2+ subgroup (from $28\%$ to $21\%$). 
If we compare the racial/ethnic discrepancy, it is the highest for the HR+/HER2- and HR-/HER2+ subgroups with respectively $15\%$ and $12\%$ differences after correction.

We see that when using our approach the estimated cure rates for black women and for white women tend to be closer than when using the KME.  This is in particular the case for the HR+/HER2+ subtype, where the cure rate for white women is almost at the same level as for black women, whereas there is a large discrepancy of $7\%$ when using the KME.  An important exception is however the subtype with the highest prevalence, namely HR+/HER2-, which still admits a high race/ethnic discrepancy of $15\%$. In conclusion, huge disparities remain for the most prevalent subtype, and although white women still enjoy better rates compared to black women, globally speaking, we observe a convergence of the cure rates between the two populations.

\vspace{.5cm}
   
\section{Proofs} \label{section_proofs}

\begin{lemma} \label{lem1}
Under the conditions of Theorem \ref{theo_incidence}, we have for any $0 \le t \le 1$, 
\begin{eqnarray}
\label{scaledprocess}
\widehat{F}_n(t\widehat{\tau}_n)=F(t\tau_c)+n^{-1/2}\Big\{(1-F(t\tau_c))Z(t\tau_c)+o_\mathbb{P}(1)\Big\},
\end{eqnarray}
where the error term is a.s.\ uniformly on $[0,1]$.
\end{lemma}

\noindent
{\bf Proof.}  
Define
\begin{eqnarray*}
Y_n(t)=\sqrt{n}\Big(\widehat{F}_n(t)-F(t)\Big),\quad t\in[0,\tau_c].
\end{eqnarray*}
Using the Skorokhod construction for $Y_n$ and the continuity of $Z$, we have the almost sure convergence of $Y_n$ to $Z$ uniformly on $[0,\tau_c]$. Next, we have for any $t\in[0,1]$,
\begin{eqnarray*}
Y_n(t\widehat{\tau}_n)-(1-F(t\tau_c))Z(t\tau_c)&=&Y_n(t\widehat{\tau}_n)-(1-F(t\widehat{\tau}_n))Z(t\widehat{\tau}_n)+(1-F(t\widehat{\tau}_n))Z(t\widehat{\tau}_n)\\
&& - (1-F(t\tau_c))Z(t\tau_c)\\
&=&(1-F(t\widehat{\tau}_n))Z(t\widehat{\tau}_n)-(1-F(t\tau_c))Z(t\tau_c)+o_\mathbb{P}(1),
\end{eqnarray*}
where $(1-F(t\widehat{\tau}_n))Z(t\widehat\tau_n)\overset{a.s}{\longrightarrow}(1-F(t\tau_c))Z(t\tau_c)$, as $n\rightarrow+\infty$, uniformly in $t\in[0,1]$ by continuity of $Z$ and $F$. Thus, we have
\begin{eqnarray*}
Y_n(.\widehat{\tau}_n)\overset{d}{\longrightarrow}(1-F(.\tau_c))Z(.\tau_c),\quad\text{as }n\rightarrow+\infty
\end{eqnarray*}
on $D[0,1]$. By differentiability of $F$, one can find a universal positive constant $K$ such that
\begin{eqnarray*}
\sqrt{n}\left|F(t\widehat{\tau}_n)-F(t\tau_c))\right|\leq K\sqrt{n}\left|\widehat{\tau}_n-\tau_c\right|.
\end{eqnarray*} 
It finally remains to prove that $\sqrt{n}\left|\widehat{\tau}_n-\tau_c\right|=o_\mathbb{P}(1)$. Note that for $\varepsilon>0$ we have that
\begin{eqnarray*}
\mathbb{P}(\sqrt{n}\left|\widehat{\tau}_n-\tau_c\right|\geq\varepsilon)\leq\mathbb{P}(\widehat{\tau}_n<\tau_c)=\Big(1-\mathbb{P}(T\geq\tau_c)\mathbb{P}(C\geq\tau_c)\Big)^n\to 0
\end{eqnarray*}
as $n\to+\infty$ since $\mathbb{P}(T\geq\tau_c)>0$ and $\mathbb{P}(C\geq\tau_c)=1-F_c(\tau_c^{-})>0$. \hfill $\Box$

\bigskip

\noindent
{\bf Proof of Theorem \ref{theo_incidence}}.  
Define
\begin{eqnarray*}
\widetilde{Z}(t):=(1-F(t\tau_c))Z(t\tau_c).
\end{eqnarray*}
In order to prove our result, we decompose $\widehat{p}_y-p_y(\tau_c)$ in the following way :

\begin{eqnarray*}
&& \widehat{p}_y-p_y(\tau_c) \\
&& = \widehat{F}_n(\widehat{\tau}_n)
+\dfrac{\widehat{F}_n(\widehat{\tau}_n)-\widehat{F}_n(y\widehat{\tau}_n)}{\widehat{y}_\gamma-1}
-F(\tau_c)+\dfrac{\left[F(\tau_c)-F(y\tau_c)\right]^2}{F(y^2\tau_c)-2F(y\tau_c)+F(\tau_c)}\\[.3cm]
&& = \widehat{F}_n(\widehat{\tau}_n)-F(\tau_c)
-\dfrac{[\widehat{F}_n(\widehat{\tau}_n)-\widehat{F}_n(y\widehat{\tau}_n)]^2}{\widehat{F}_n(y^2\widehat{\tau}_n)-2\widehat{F}_n(y\widehat{\tau}_n)+\widehat{F}_n(\widehat{\tau}_n)}
+\dfrac{\left[F(\tau_c)-F(y\tau_c)\right]^2}{F(y^2\tau_c)-2F(y\tau_c)+F(\tau_c)}\\[.3cm]
&& = \widehat{F}_n(\widehat{\tau}_n)-F(\tau_c)
-\dfrac{[\widehat{F}_n(\widehat{\tau}_n)-\widehat{F}_n(y\widehat{\tau}_n)]^2-\left[F(\tau_c)-F(y\tau_c)\right]^2}{F(y^2\tau_c)-2F(y\tau_c)+F(\tau_c)}\\
&& \hspace*{.5cm} - \left[\dfrac{1}{\widehat{F}_n(y^2\widehat{\tau}_n)-2\widehat{F}_n(y\widehat{\tau}_n)+\widehat{F}_n(\widehat{\tau}_n)}-\dfrac{1}{F(y^2\tau_c)-2F(y\tau_c)+F(\tau_c)}\right][\widehat{F}_n(\widehat{\tau}_n)-\widehat{F}_n(y\widehat{\tau}_n)]^2\\[.3cm]
&& = A_1+A_2+A_3.
\end{eqnarray*}

The term $A_1$ can be handled by using the decomposition given in Lemma \ref{lem1} for $t=1$~:
\begin{eqnarray*}
A_1 = n^{-1/2} (\widetilde{Z}(1)+o_\mathbb{P}(1)).
\end{eqnarray*}

For $A_2$ and $A_3$, we respectively use a Taylor expansion of the square and the inverse function, combined with an application of Lemma \ref{lem1}. This gives
\begin{eqnarray*}
A_2&=&-2\left[\widehat{F}_n(\widehat{\tau}_n)-\widehat{F}_n(y\widehat{\tau}_n)-F(\tau_c)+F(y\tau_c)\right]\dfrac{F(\tau_c)-F(y\tau_c)}{F(y^2\tau_c)-2F(y\tau_c)+F(\tau_c)}(1+o_\mathbb{P}(1))\\[.3cm]
&=&-2n^{-1/2}\left[\widetilde{Z}(1)-\widetilde{Z}(y)+o_\mathbb{P}(1)\right]\dfrac{F(\tau_c)-F(y\tau_c)}{F(y^2\tau_c)-2F(y\tau_c)+F(\tau_c)}(1+o_\mathbb{P}(1))
\end{eqnarray*}
and
\begin{eqnarray*}
A_3&=&\dfrac{\widehat{F}_n(y^2\widehat{\tau}_n)-2\widehat{F}_n(y\widehat{\tau}_n)+\widehat{F}_n(\widehat{\tau}_n)-F(y^2\tau_c)+2F(y\tau_c)-F(\tau_c)}{[F(y^2\tau_c)-2F(y\tau_c)+F(\tau_c)]^2}(1+o_\mathbb{P}(1)) \\
&& \times [F(\tau_c)-F(y\tau_c)+o_\mathbb{P}(1)]^2\\[.3cm]
&=&n^{-1/2}\left[\widetilde{Z}(y^2)-2\widetilde{Z}(y)+\widetilde{Z}(1)+o_\mathbb{P}(1)\right]\left[\dfrac{F(\tau_c)-F(y\tau_c)}{F(y^2\tau_c)-2F(y\tau_c)+F(\tau_c)}\right]^2(1+o_\mathbb{P}(1)).
\end{eqnarray*}

By direct summation, we obtain
\begin{eqnarray*}
\widehat{p}_y-p_y(\tau_c)&=&n^{-1/2}\left[\sum_{i=0}^2a_i(\tau_c)\widetilde{Z}(y^i)+o_\mathbb{P}(1)\right],
\end{eqnarray*}
with
\begin{eqnarray*}
a_0(\tau_c)&=&\left[1-\dfrac{F(\tau_c)-F(y\tau_c)}{F(y^2\tau_c)-2F(y\tau_c)+F(\tau_c)}\right]^2,\\
a_1(\tau_c)&=&-2\dfrac{F(\tau_c)-F(y\tau_c)}{F(y^2\tau_c)-2F(y\tau_c)+F(\tau_c)}\left[1+\dfrac{F(\tau_c)-F(y\tau_c)}{F(y^2\tau_c)-2F(y\tau_c)+F(\tau_c)}\right],\\
a_2(\tau_c)&=&\left[\dfrac{F(\tau_c)-F(y\tau_c)}{F(y^2\tau_c)-2F(y\tau_c)+F(\tau_c)}\right]^2.
\end{eqnarray*}

Hence, the asymptotic variance $\sigma^2(\tau_c)$ is given by
\begin{eqnarray*}
\sigma^2_{y,\tau_c}=\sum_{i,j=0}^2a_i(\tau_c)a_j(\tau_c)\text{cov}\Big(\widetilde{Z}(y^i),\widetilde{Z}(y^j)\Big),
\end{eqnarray*}
with
\begin{eqnarray*}
\text{cov}\Big(\widetilde{Z}(y^i),\widetilde{Z}(y^j)\Big)=(1-F(y^i\tau_c))(1-F(y^j\tau_c))v(y^{i\vee j}\tau_c). 
\end{eqnarray*}
\hfill $\Box$

\newpage

\bibliographystyle{dcu}
\bibliography{bibli.bib}

\end{document}